\documentclass[iop]{emulateapj}

\newcommand{\xmm} {{\it XMM-Newton}}
\newcommand{\chandra} {{\it Chandra}}
\newcommand{\nustar} {{\it NuSTAR}}
\newcommand{\suzaku} {{\it Suzaku}}
\newcommand{\swift} {{\it Swift}}
\newcommand{\swiftxrt} {{\it Swift}/XRT}

\newcommand{\cmsq} {cm$^{-2}$}
\newcommand{\nh} {$N_{\rm{H}}$}
\newcommand{\lx} {$L_{\rm{X}}$}
\newcommand{\fx} {$F_{\rm{X}}$}
\newcommand{\chisq} {$\chi^2$}
\newcommand{\dchisq} {$\Delta\chi^2$}
\newcommand{\rchisq} {$\chi^2_r$}

\newcommand{\degree}{{$^\circ$}}
\newcommand{\countss}{\mbox{\thinspace counts\thinspace s$^{-1}$}}
\newcommand{\ergs}{\mbox{\thinspace erg\thinspace s$^{-1}$}}
\newcommand{\ergcms}{\mbox{\thinspace erg\thinspace cm$^{-2}$\thinspace s$^{-1}$}}

\newcommand{\mbh} {$M_{\rm BH}$}

\newcommand{\lamedd} {$\lambda_{\rm Edd}$}

\newcommand{\msol} {$M_{\odot}$}

\shorttitle{.}
\shortauthors{Brightman et al.}

\begin{document}

\title{A broadband X-ray spectral study of the intermediate-mass black hole candidate M82~X-1 with NuSTAR, Chandra and Swift}

\author{Murray Brightman$^{1}$, Fiona A. Harrison$^{1}$, Didier Barret$^{2,3}$, Shane W. Davis$^{4}$, Felix F{\"u}rst$^{1}$, Kristin K. Madsen$^{1}$, Matthew Middleton$^{5}$, Jon M. Miller$^{6}$, Daniel Stern$^{7}$, Lian Tao$^{1}$, Dominic J. Walton$^{7,1}$}

\affil{$^{1}$Cahill Center for Astrophysics, California Institute of Technology, 1216 East California Boulevard, Pasadena, CA 91125, USA\\
$^{2}$Universite de Toulouse, UPS-OMP, IRAP, Toulouse, France\\
$^{3}$CNRS, IRAP, 9 Av. colonel Roche, BP 44346, F-31028 Toulouse cedex 4, France\\
$^{4}$Department of Astronomy, University of Virginia, P.O. Box 400325, Charlottesville, VA 22904-4325, USA\\
$^{5}$Institute of Astronomy, Madingley Road, Cambridge CB3 OHA, UK\\
$^{6}$Department of Astronomy, University of Michigan, 1085 S. University Ave, Ann Arbor, MI 48109-1107, USA\\
$^{7}$Jet Propulsion Laboratory, California Institute of Technology, Pasadena, CA 91109, USA\\
}

\begin{abstract}

M82~X-1 is one of the brightest ultraluminous X-ray sources (ULXs) known, which, assuming Eddington-limited accretion and other considerations, makes it one of the best intermediate-mass black hole (IMBH) candidates. However, the ULX may still be explained by super-Eddington accretion onto a stellar-remnant black hole. We present simultaneous \nustar, \chandra\ and \swiftxrt\ observations during the peak of a flaring episode with the aim of modeling the emission of M82~X-1 and yielding insights into its nature. We find that thin-accretion disk models all require accretion rates at or above the Eddington limit in order to reproduce the spectral shape, given a range of black hole masses and spins. Since at these high Eddington ratios the thin-disk model breaks down due to radial advection in the disk, we discard the results of the thin-disk models as unphysical. We find that the temperature profile as a function of disk radius ($T(r)\propto r^{-p}$) is significantly flatter ($p=0.55^{+ 0.07}_{- 0.04}$) than expected for a standard thin disk ($p=0.75$). A flatter profile is instead characteristic of a slim disk which is highly suggestive of super-Eddington accretion. Furthermore, radiation hydrodynamical simulations of super-Eddington accretion have shown that the predicted spectra of these systems are very similar to what we observe for M82~X-1. We therefore conclude that M82~X-1 is a super-Eddington accretor. Our mass estimates inferred from the inner disk radius imply a stellar-remnant black hole (\mbh=$26^{+9}_{-6}$~\msol) when assuming zero spin, or an IMBH (\mbh=$125^{+45}_{-30}$~\msol) when assuming maximal spin.

\end{abstract}


\section{Introduction}

The ultraluminous X-ray source M82~X-1 is one of the best candidates for an intermediate-mass black hole ($100<M_{\rm BH}<10000$~\msol) based on several indirect factors. These include the source's high luminosity, which can reach $\sim10^{41}$~\ergs\ \citep[e.g.][]{ptak99b, rephaeli02, kaaret06}, far greater than the Eddington limit of a stellar-remnant black hole of mass $\sim$10 \msol\ that is typical of X-ray binaries in our own Galaxy ($\sim10^{39}$~\ergs); detection of low-frequency quasi-periodic oscillations (QPOs) in the power spectrum \citep[54 mHz,][]{strohmayer03, dewangan06, mucciarelli06}, indicative of a compact, unbeamed source; as well as twin-peaked QPOs at 3.3 and 5.1 Hz, which lead to a mass estimate using scaling laws between the QPOs frequencies and mass \citep{pasham14}. The mass estimates for X-1 vary considerably, however, and have large uncertainties. This makes its status as an IMBH is not yet firmly established. The most recent estimate came from the twin peak QPOs, which give a mass of 428$\pm$105~\msol\ \citep{pasham14}. On the other hand, modeling of the accretion disk emission by \cite{okajima06} instead found that the source can be explained by a $\sim30$~\msol\ stellar remnant black hole radiating at several times its Eddington limit.

At moderate Eddington ratios (\lamedd$\equiv L/L_{\rm Edd}\ll1$), the accretion on to a black hole can be described by the standard ``thin'' accretion disk model \citep[][SS73]{shakura73}. For the standard disk model, the accretion disk is geometrically thin and optically thick where viscous heating in the disk is balanced by radiative cooling and the local temperature of the disk, $T$, decreases with radius, $r$ as $T(r)\propto r^{-0.75}$. Under the assumption that the disk extends down to the innermost stable circular orbit \citep[ISCO, e.g.][]{steiner10}, spectral modeling yields the temperature of the disk at the ISCO, which in turn yields the inner radius. The inner radius is directly proportional to the mass of the black hole, albeit with a large degeneracy with the black hole's spin, which can be used for mass estimates. 

However, as the mass accretion rate increases, advective cooling dominates over radiative cooling and the thin-disk model breaks down. The scale height of the disk increases and thus is referred to as the ``slim'' disk model \citep{abramowicz88}. For a slim disk, the local temperature of the disk has a flatter temperature profile as a function of radius with $T(r)\propto r^{-0.5}$ \citep{watarai00}. Slim disks have been proposed as mechanisms to explain ULXs as super-Eddington stellar remnant black hole accretors rather than IMBHs \citep[e.g.][]{kato98,poutanen07}. In addition to the modified disk spectrum, the emission from super-Eddington accretion is expected to produce winds/outflows \citep{king03} which may also modify the emission spectrum \citep[e.g. via Compton scattering,][]{kawashima09}. Indeed, high-velocity, ionized outflows have recently been detected in the high-resolution X-ray grating spectra of NGC~1313~X-1 and NGC~5408~X-1 \citep{pinto16} and confirmed in a follow-up study with CCD resolution data for NGC~1313~X-1 \citep{walton16}.

Therefore, spectral modeling of the emission from ULXs and testing for a departure from the thin-disk model can yield important information regarding their nature. For M82~X-1, however, modeling of the disk emission has yielded conflicting results. \cite{feng10} observed the source with \xmm\ and \chandra\ over the course of a flaring episode and fitted the spectra with the standard thin-disk model. They found that the luminosity of the disk, $L$, scaled with inner temperature as $L\propto T^4$ which is expected from a thin accretion disk with a constant inner radius. From this they inferred a black hole mass in the range $300-810$~\msol, assuming that the black hole is rapidly spinning in order to avoid extreme violations of the Eddington limit.

However, using a different \xmm\ dataset, \cite{okajima06} modeled the emission of the accretion disk from M82~X-1 instead finding that the temperature profile of the disk was too flat ($T(r)\propto r^{0.61}$) to be consistent with the standard thin-accretion disk and concluded that it was in the slim-disk condition. Applying a theoretical slim-disk model their estimate for the mass of the black hole was \mbh$\approx19-32$\msol. Considering broader band data afforded by \suzaku/XIS and HXD-PIN allowed \cite{miyawaki09} (M09) to better distinguish between thin and slim-disk models. While M09 also consider a slim-disk conclusion based on a high inferred Eddington ratio, they instead prefer the power-law state interpretation, finding the spectrum to be too hard to be explained by emission from an optically thick accretion disk.

These conflicting results may stem from the fact that spectral studies of X-1 are complicated by the presence of another ultraluminous X-ray source only 5\arcsec\ from X-1, which was recently identified as an ultraluminous X-ray pulsar \citep{bachetti14}. Since this source can reach luminosities of $10^{40}$~\ergs\ \citep{feng07, kong07,brightman16}, and is only resolvable from X-1 with \chandra, its contribution to the X-ray spectrum of M82 must be taken into account when modeling the spectrum of X-1 with other X-ray instruments. Furthermore, X-1 and X-2 are embedded in bright diffuse emission \citep{ranalli08}, which further complicates analysis. X-1 is also bright enough to cause pile-up effects on the \chandra\ detectors, which can severely distort the spectrum. 

In this paper we report on simultaneous observations of M82 with \nustar, \chandra\ and \swiftxrt\ during an episode of flaring activity from X-1. We aim to improve upon previous works with the combination of \chandra\ to spatially resolve X-1 from X-2 and the diffuse emission below 8 keV, and  \nustar\ to gain broadband sensitive spectral coverage, especially above 10 keV. Our goal is to determine if the emission from the disk is indeed consistent with a standard thin-accretion disk, which would support the IMBH scenario, or if it shows a significant departure from this model that would indicate a super-Eddington accretor of lower mass.

In section \ref{sec_obs} we describe our observations, including details of the \swiftxrt\ monitoring that showed the increased flux from M82, triggering our \nustar\ and \chandra\ Director's Discretionary Time (DDT) requests, including details of the data reduction, while in section \ref{sec_spec} we describe our spectral analysis where we test various emission models for X-1. In section \ref{sec_x1} we describe the results from the disk models and the mass estimation of X-1. In section \ref{sec_comp} we discuss our results with respect to previous analyses and we finish with a discussion of alternative interpretations of the high-energy spectrum section \ref{sec_alt}. We conclude and summarize in section \ref{sec_conc}. A distance of 3.3 Mpc to M82 is assumed throughout \citep{foley14}.

\section{Observations and Data reduction}
\label{sec_obs}

The \chandra\ and \nustar\ observations were taken simultaneously, along with a \swift/XRT observation between 2015 June 20$-$21. Table \ref{table_obsdat} provides a description of the observational data. The following sections describe the individual observations and data reduction. Spectral fitting was carried out using {\sc xspec} v12.8.2 \citep{arnaud96} and all uncertainties quoted are at the 90\% level.

\begin{table*}
\centering
\caption{Observational data}
\label{table_obsdat}
\begin{center}
\begin{tabular}{l c c c c r c l}
\hline
Observatory	& ObsID	& Start date (UT)	& End date (UT)	& Instrument	& Exposure	& Energy band	& Count rate \\
			&		&			& 			&			& (ks) 		& (keV)		& (counts s$^{-1}$) \\
%(1) & (2) & (3) & (4) & (5) & (6)  \\
\hline
\nustar\	& 90101005002	& 2015-06-20T14:21:07	& 2015-06-21T06:21:07	& FPMA	& 37.4 	& 3$-$30	& 0.88 \\
		&				&					&					& FPMB	& 37.4	& 3$-$30	& 0.87 \\
\swift\	& 00081460001	& 2015-06-20T17:33:18	& 2015-06-20T19:21:55	& XRT	& 1.6		& 0.5$-$10& 0.77 \\
\chandra\	& 17678			& 2015-06-21T02:46:16	& 2015-06-21T06:17:39	& ACIS-I	&  9.3	& 0.5$-$8	& 2.45 \\
 \hline
\end{tabular}
\end{center}
\end{table*}

\subsection{Swift/XRT}

\begin{figure*}
\begin{center}
\includegraphics[width=180mm]{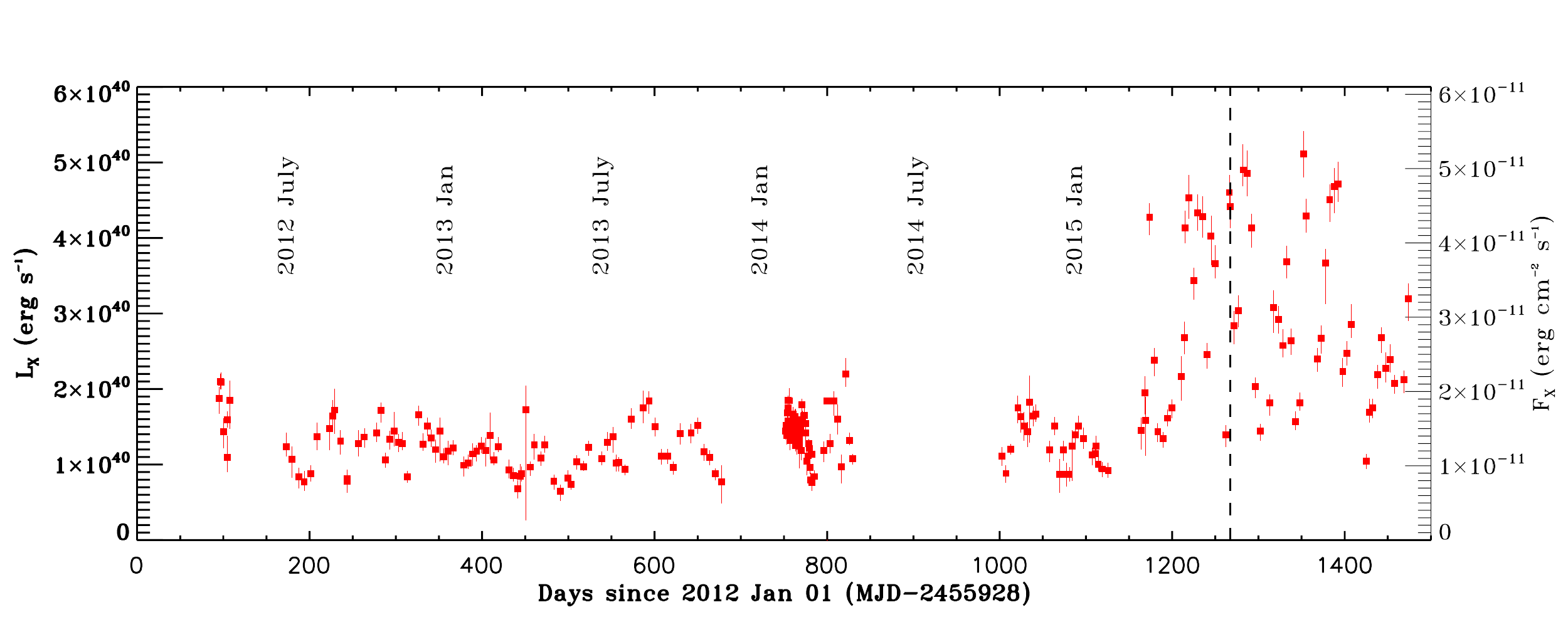}
\caption{\swift/XRT 2$-$10~keV light curve of M82 from 2012 April 05 until 2016 January 14. 2015 June 21, the time at which the joint \nustar, \chandra\ and \swiftxrt\ observations were taken, is indicated with a dashed line, during which the emission from M82 was at its highest point since \swift/XRT monitoring had begun.}
\label{fig_swift_ltcrv}
\end{center}
\end{figure*}

M82 has been monitored by \swiftxrt\ \citep{burrows05} every few days since 2012 (albeit with several gaps), including a period of intense monitoring of SN2014J at the beginning of 2014. Figure \ref{fig_swift_ltcrv} shows a light curve of the 2$-$10~keV emission from the galaxy, which was produced from all 238 \swift/XRT observations of the galaxy taken since the beginning of 2012 until the beginning of 2016. We used the {\sc heasoft} tool {\sc xselect} to filter events from a 49\arcsec\ radius region centered on the peak of the emission and to extract the spectrum. Background events were extracted from a nearby circular region of the same size. The spectra were grouped with a minimum of 1 count per bin and the Cash statistic was used for spectral fitting. The spectra were fitted in the range 0.2$-$10~keV with an absorbed power-law. The luminosity in Figure \ref{fig_swift_ltcrv} is the observed luminosity between 2$-$10~keV given a redshift of 0.00067. The observed flux in the 2$-$10~keV band is shown on the right axis.

As shown in Figure \ref{fig_swift_ltcrv}, between 2012 and early 2015 the observed 2$-$10~keV flux from the galaxy varied between 1$-$2$\times10^{-11}$~\ergcms, implying a 2$-$10~keV luminosity of 1$-$2$\times10^{40}$~\ergs. On 2015 March 20 the flux from the galaxy increased by several factors to $>4\times10^{-11}$~\ergcms, and continued to flare up to this level for several months. The \swiftxrt\ images pointed towards one of the ULXs as the origin of the brightening, however its spatial resolution and pointing accuracy did not allow us to pinpoint the emission.

\subsection{Chandra}
\begin{figure}
\begin{center}
\includegraphics[width=80mm]{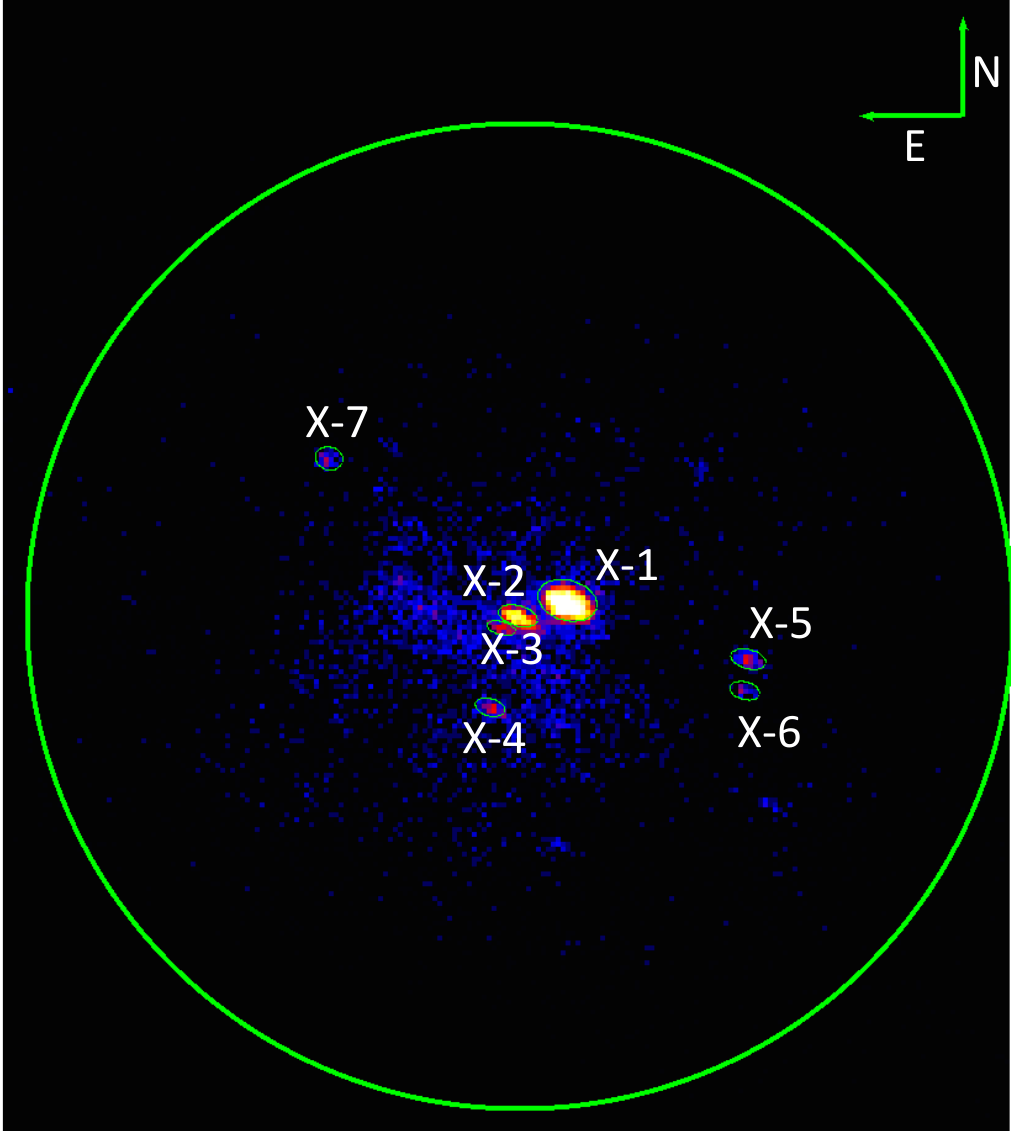}
\caption{\chandra\ ACIS-I 0.5$-$8~keV image of M82 taken on 2015 July 21. The large green circle shows the \nustar\ and \swift/XRT extraction region which has a radius of 49\arcsec. The brightest point-sources within this region are labelled. North is up, east is left, indicated by the arrows in the upper right corner, which are 10\arcsec\ long. }
\label{fig_chandra_img}
\end{center}
\end{figure}

Since the angular separation of X-1 and X-2 is only 5\arcsec, only \chandra\ \citep{weisskopf99} can spatially resolve the emission from these two sources. We were therefore granted a DDT \chandra\ observation to ascertain the origin of the increased X-ray flux from M82. This observation was taken with ACIS-I at the optical axis with only a 1/8th sub-array of pixels on chip I3 turned on. M82 was placed 4\arcmin\ off-axis to smear out the PSF in order to mitigate the effect of pile-up from the bright point-sources. The sub-array of pixels was used to decrease the readout time of the detector, further mitigating the effect of pile-up. Figure \ref{fig_chandra_img} shows the \chandra\ image of the central region of M82, which shows that the cause of the increased X-ray emission from M82 was X-1. We also indicate the other point-sources in this figure, including the ultraluminous pulsar (X-2). 

We proceeded to extract the \chandra\ spectra of the point-sources using the {\sc ciao} (v4.7, CALDB v4.6.5) tool {\sc specextract}, from elliptical regions to encompass the shape of the off-axis \chandra\ PSF. For X-1 we used a semi-major axis of 3\arcsec\ and a semi-minor axis of 2\arcsec. For X-2 we used 2\arcsec\ and 1\arcsec\ and for the other sources between 1$-$2\arcsec. A small region close to the center of the image was used for background subtraction for X-1, X-2 and X-3. For the other sources, a background region outside the galaxy was used. 

\begin{figure}
\begin{center}
\includegraphics[width=90mm]{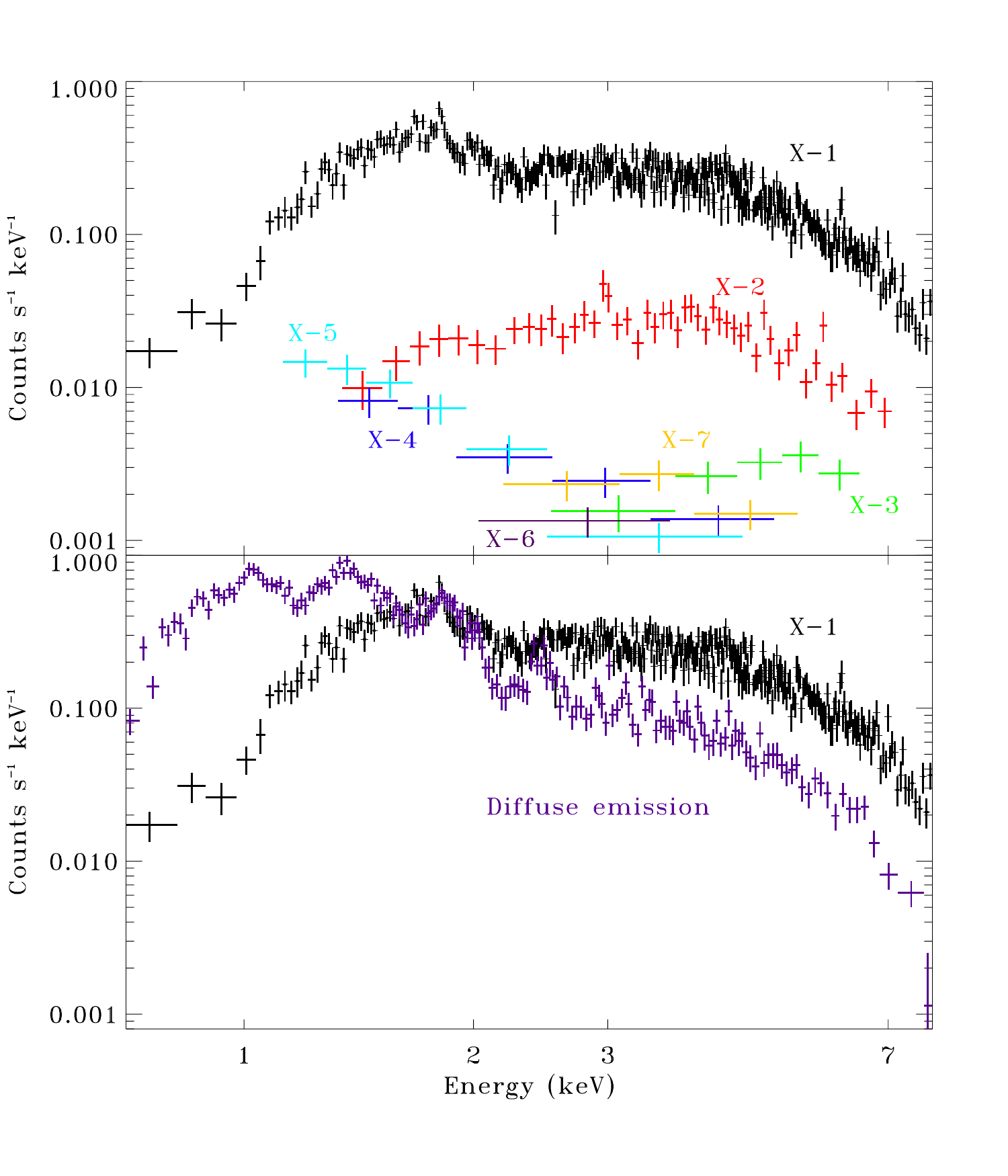}
\caption{\chandra\ ACIS-I spectra of the seven brightest point-sources within the \nustar\ and \swift/XRT extraction regions (top) and the spectrum of the diffuse emission in M82 within the same region compared to the brightest source, X-1 (bottom). }
\label{fig_chandra_spec}
\end{center}
\end{figure}

Figure \ref{fig_chandra_spec} shows the \chandra\ spectra of all these sources, including the spectrum of the diffuse emission within the central 49\arcsec\ of the galaxy. X-1 dominates the emission from the galaxy in the \chandra\ band with 1.29~\countss, whereas the diffuse emission contributes 1.03~\countss, X-2 contributes 0.11~\countss\ and the other point sources contribute $\leq$0.01~\countss\ individually. 

\subsection{NuSTAR}
In addition to the \chandra\ observation, we were granted a \nustar\ \citep{harrison13} DDT observation to obtain a source-dominated spectrum of M82 above 10~keV during a flaring event of X-1. 

The raw \nustar\ data were reduced using the {\sc nustardas} software package version 1.4.1. The events were cleaned and filtered using the {\tt nupipeline} script with standard parameters. The {\tt nuproducts} task was used to generate the spectra and the corresponding response files. Spectra were extracted from a circular aperture of radius 49\arcsec\ centered on the peak of the emission shown in Figure \ref{fig_chandra_img}. The background spectra were extracted from a region encompassing the same detector chip as the source, excluding the source extraction region and avoiding the wings of the PSF as much as possible.

\begin{figure}
\begin{center}
\includegraphics[width=90mm]{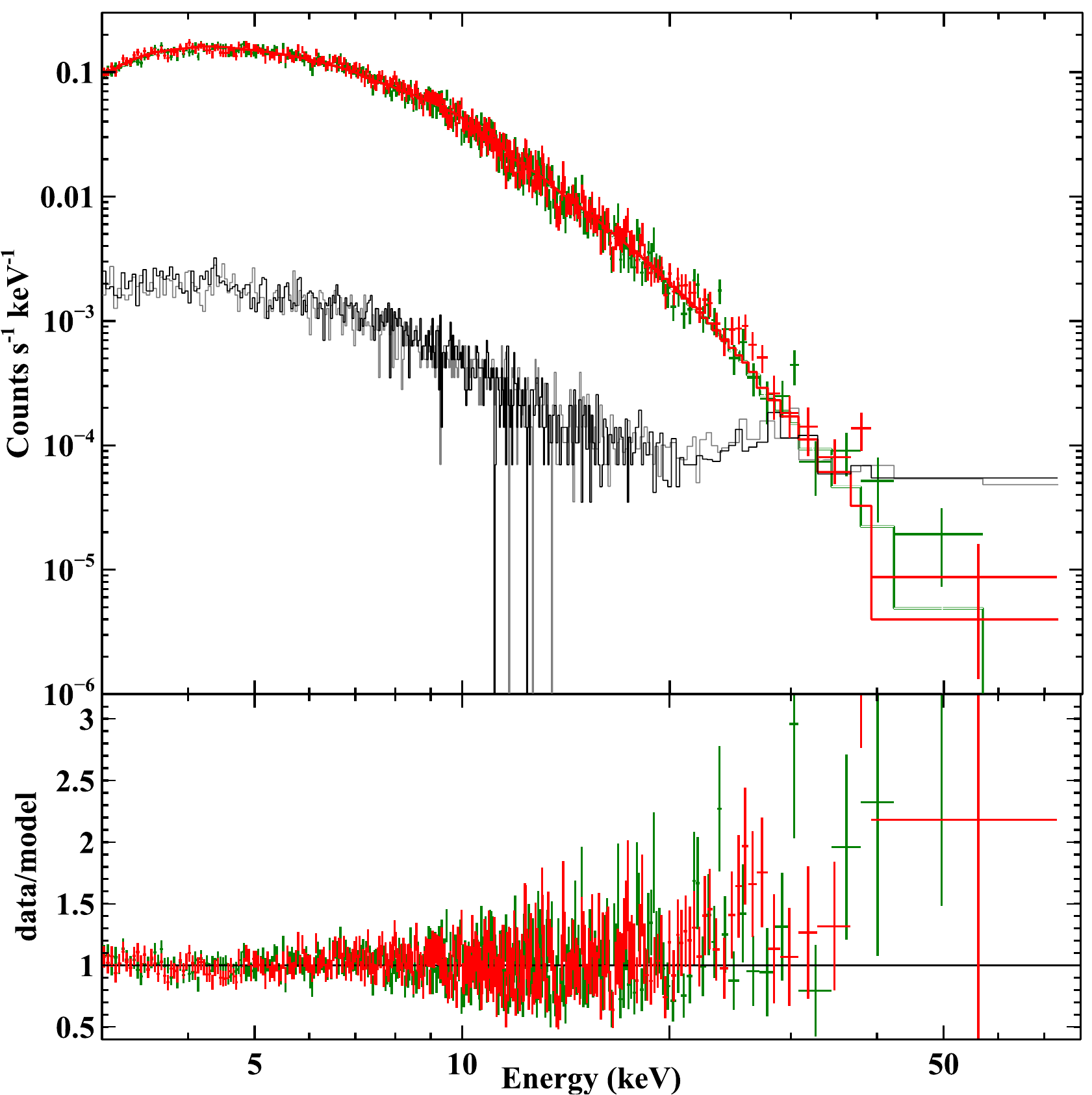}
\caption{\nustar\ 3$-$79~keV FPMA (red) and FPMB (green) spectra of M82 taken on 2015 July 20-21 (top) which are fitted with an absorbed cutoff power-law model. The backgrounds are plotted as black and grey histograms respectively. The data are source-dominated between 3$-$30~keV, above which the data are dominated by the background. The bottom panel shows the data-to-model ratio of the fit with the absorbed cutoff power-law model.}
\label{fig_nustar_spec}
\end{center}
\end{figure}

Data from both focal plane modules (FPMA and FPMB) are used for simultaneous fitting, without co-adding. Figure \ref{fig_nustar_spec} shows the \nustar\ spectrum of M82 from this observation. The background is also plotted showing that the spectrum remains source-dominated up to $\sim$30~keV. For this reason, we limit our spectral analysis to below 30~keV. The \nustar\ spectrum can be described with an absorbed cutoff power-law ({\tt zwabs*cutoffpl}) where \nh$=3.7\pm0.7\times10^{22}$ \cmsq, $\Gamma=1.74^{+0.15}_{-0.16}$ and $E_{\rm C}=8.89^{+1.30}_{-1.05}$~keV  with a normalization of $2.894^{+0.005}_{-0.004}\times10^{-2}$, which is shown in Figure \ref{fig_nustar_spec}. The absorption model {\tt zwabs} uses Wisconsin \citep{morrison83} cross-sections and \cite{anders82} abundances. We leave the cross-normalization between the two FPMs to float, where $C_{\rm B}/C_{\rm A}=1.05\pm0.01$, which is close to typical values \citep{madsen15}. The flux in the $3-79$~keV band is 5.50$\times10^{-11}$~\ergcms.

\section{Spectral analysis and results}
\label{sec_spec}

\subsection{Constraining the diffuse emission}
\label{sec_chan_spec}

While X-1 dominates the X-ray emission from M82 during our observations, the diffuse emission and X-2 contribute significantly to the flux in the \chandra\ band within the \nustar\ and \swift\ extraction regions. Therefore in order to properly model the broadband emission from X-1, we must account for these other significant sources of emission. For our spectral analysis we model the fainter point sources as part of the diffuse emission. 

\begin{figure}
\begin{center}
\includegraphics[width=90mm]{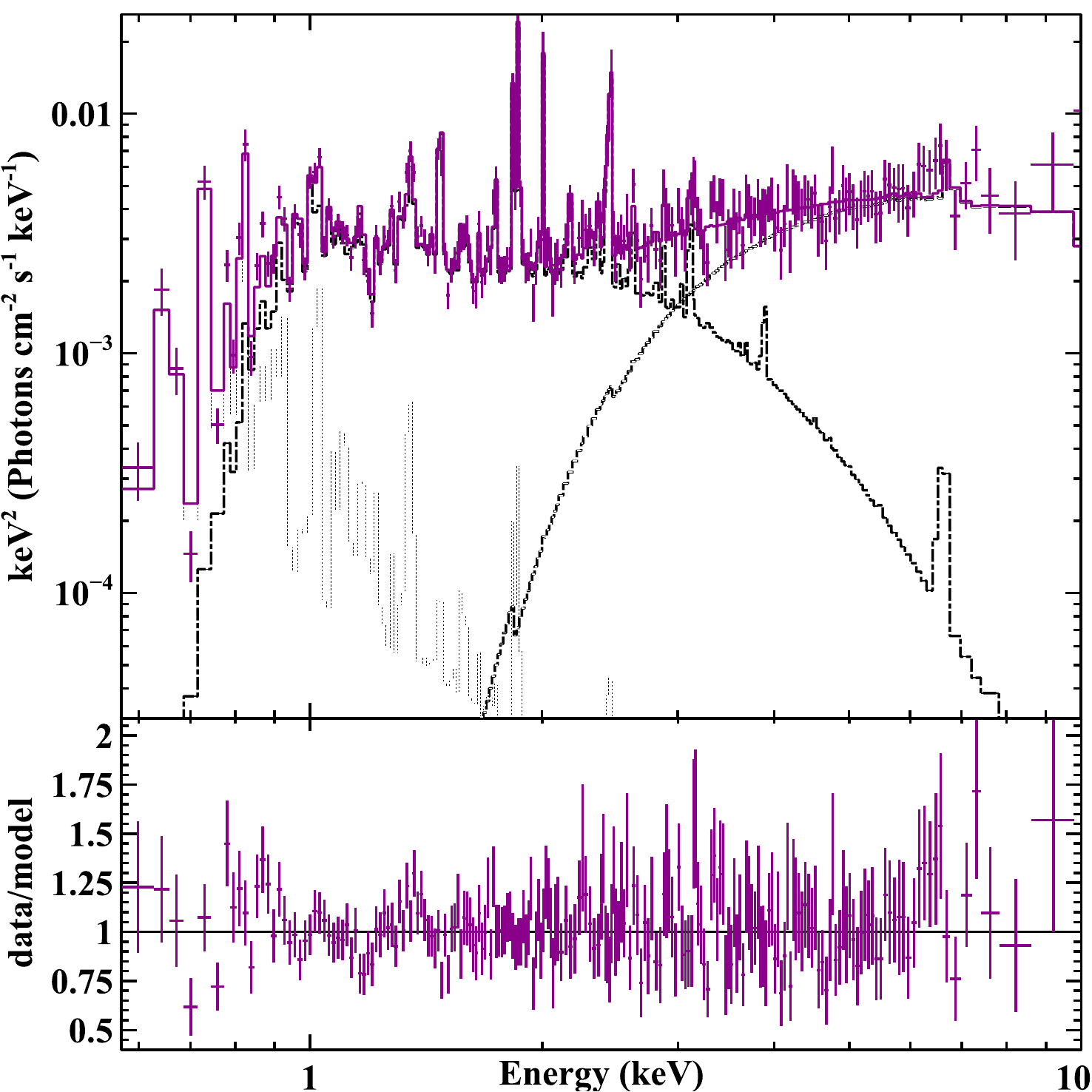}
\caption{\chandra\ 0.5$-$8~keV ACIS-I spectrum of the diffuse emission from M82 extracted from the same 49\arcsec\ region used for the \nustar\ and \swift/XRT spectral extraction, but with the emission from X-1 and X-2 masked out. The top panel shows the unfolded spectrum fitted with a combination of three {\sc apec} models subjected to line-of-sight photo-electric absorption. The bottom panel shows the data-to-model ratio.}
\label{fig_chandra_diffuse}
\end{center}
\end{figure}

The diffuse emission from M82 was studied in depth by \cite{ranalli08} using deep \xmm\ EPIC and RGS data. They find that this emission is best described by thermal plasma emission from hot gas with a double-peaked temperature, with peaks at $\sim0.5$ and $\sim7$~keV, for which they use the {\tt apec} model in {\sc xspec}. Following this we fit the diffuse emission with a combination of {\tt apec} models, subjected to photo-electric absorption. We leave the abundances to vary for each component as well as the \nh. We find that the $E<1$~keV spectrum requires two {\tt apec} models with temperatures of $\sim0.3$ and $\sim1$~keV. We note that the temperature distribution of the lower energy peak in \cite{ranalli08} is broad, which is most likely why we find the need for two components rather than one. For the third high-energy component, we find that the temperature is not well constrained, so we fix this value at 7~keV. The spectral fit of the diffuse emission is shown in Figure \ref{fig_chandra_diffuse} and the spectral parameters are given in Table \ref{table_diffpar}. The \chisq/degrees of freedom (DOF) of this fit is 172.64/204=0.846. The 0.5$-$8~keV flux (luminosity) of the diffuse emission is 1.47$\times10^{-11}$~\ergcms\ (1.91$\times10^{40}$~\ergs) during this observation.

\begin{table}
\centering
\caption{Spectral parameters of the diffuse emission}
\label{table_diffpar}
\begin{center}
\begin{tabular}{l l l l}
\hline
					& {\tt apec 1}				& {\tt apec 2}				& {\tt apec 3}		\\
\hline
\nh\ ($10^{22}$\cmsq)	& 1.45$^{+0.17}_{-0.10}$		& 0.74$^{+0.13}_{-0.17}$		& 9.28$^{+2.24}_{-1.76}$ \\
$kT$	(keV)			& 1.01$\pm0.05$			& 0.28$\pm0.03$			& 7 (fixed)					\\
Adund.				& 0.78$^{+0.26}_{-0.57}$		& 4.96$^{+u}_{-4.27}$		& 0.04$^{+0.13}_{-0.04}$	\\
norm	.				& 0.024$^{+0.007}_{-0.006}$	& 0.006$^{+0.04}_{-0.001}$	& 0.013$\pm0.002$ \\
\hline
\end{tabular}
\end{center}
\end{table}

\subsection{Accounting for emission from X-2}

A detailed spectral analysis of the \chandra\ spectra of X-2 was presented in \cite{brightman16} which found that these could be equally well described by absorbed power-law, cutoff power-law or disk blackbody models. While only \chandra\ can spatially resolve X-2 from nearby point-sources, its limited bandpass does not allow it to constrain broad band emission models for X-2. \nustar\ on the other hand cannot spatially resolve X-2. However, due to the coherent pulsations from the source and the timing capabilities of \nustar, the pulsed emission in the 3$-$79~keV band was measured. This pulsed spectrum is well described by an absorbed (\nh\ fixed to $3\times10^{22}$ \cmsq) cutoff power-law ({\tt cutoffpl}), where $\Gamma=0.6\pm0.3$ and $E_{\rm C}=14^{+5}_{-3}$~keV. The pulsed spectrum compares well with the phase-averaged \chandra\ spectra using the same model, which yields $\Gamma=0.70^{+0.68}_{-0.65}$ and $E_{\rm C}=6.19^{+50.9}_{-2.9}$~keV. Therefore in our present analysis, we model the emission from X-2 with the {\tt cutoffpl} model. From our new \chandra\ data on X-2, we measure \nh=2.92$^{+0.73}_{-0.85}\times10^{22}$ \cmsq, $\Gamma=1.16^{+0.44}_{-1.70}$ and an unconstrained cutoff energy. The 0.5$-$8~keV flux (luminosity) from X-2 was 5.94$\times10^{-12}$~\ergcms\ (7.73$\times10^{39}$~\ergs) during our observation.

\subsection{Combined spectral analysis of X-1}

Finally, for X-1, we find that the \chandra\ spectra are heavily piled-up with a pile-up fraction of $>10$\%, despite our efforts to mitigate this with a sub-array of pixels and by locating the source off axis. For this reason we opt not to use the \chandra\ data on X-1 and instead infer the spectral properties of X-1 from the \swift/XRT and \nustar\ data, having accounted for the diffuse emission and X-2 with modeling of the \chandra\ data.

For our broadband spectral study of X-1 we conduct simultaneous fitting of the combined \chandra, \nustar\ and \swiftxrt\ datasets. In Section \ref{sec_chan_spec} we described our spectral fitting of the diffuse emission within the \nustar\ and \swiftxrt\ extraction regions. We fix all the spectral parameters of this component in our joint fit. For X-2, modeled with a cutoff power-law, we allow the spectral parameters to vary in the fit, since X-2 is expected to contribute significantly above 8~keV and be constrained by the \nustar\ data. For X-1 we test a variety of emission models.

We use five spectral datasets in our joint fit, the \chandra\ 0.5$-$8~keV spectrum of the diffuse emission, the \chandra\ 0.5$-$8~keV spectrum of X-2, the \swiftxrt\ 0.5$-$10~keV spectrum of the integrated emission from M82 and the \nustar\ FPMA and FPMB 3$-$79~keV spectrum of the integrated emission from M82. For each dataset we allow a cross-normalization constant to vary to allow for instrument cross-calibration uncertainties that cannot vary beyond $\pm10$\% of the FPMA normalization \citep{madsen15}.  

\section{Results from the disk models for X-1}
\label{sec_x1}

\begin{figure*}
\begin{center}
\includegraphics[width=160mm]{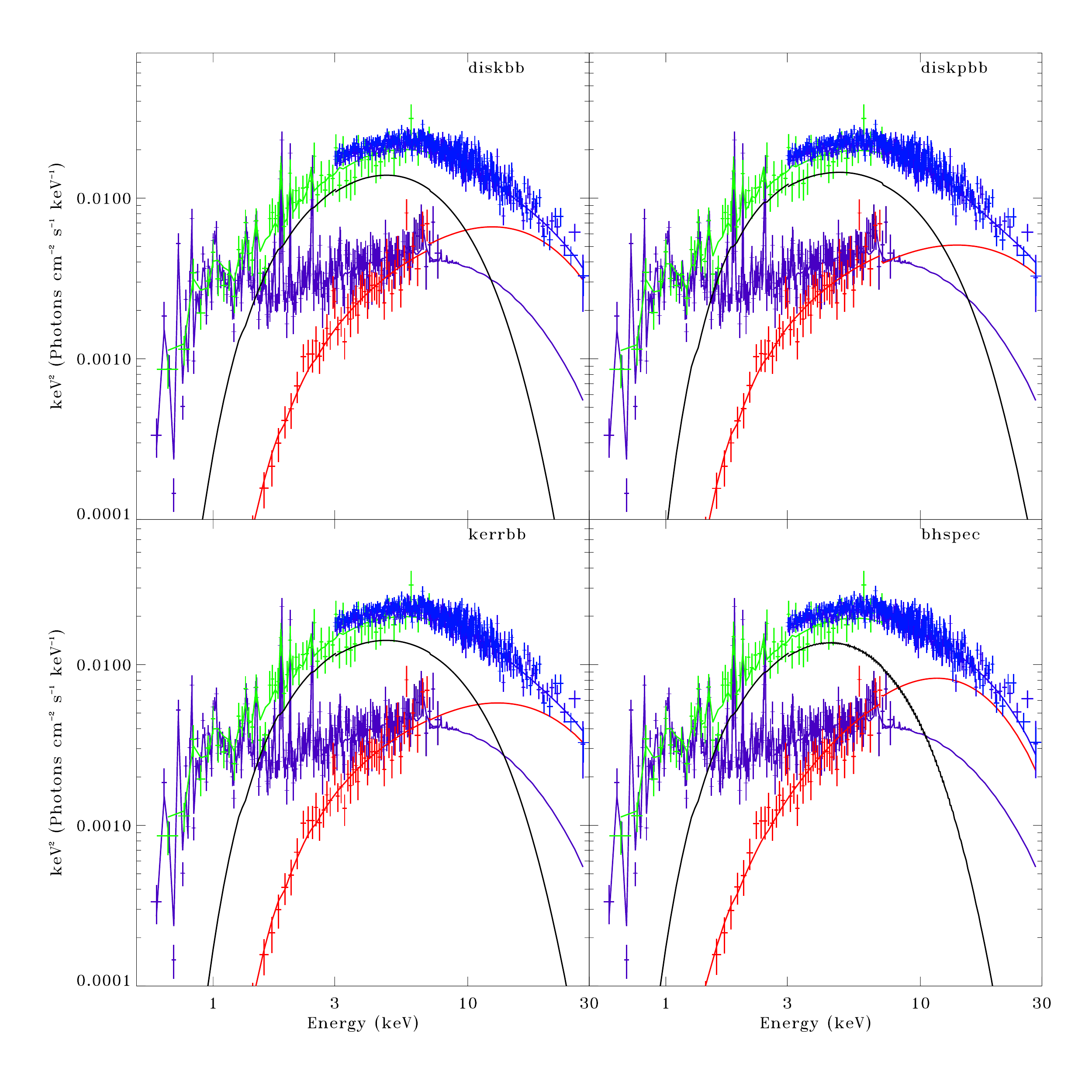}
\includegraphics[width=160mm]{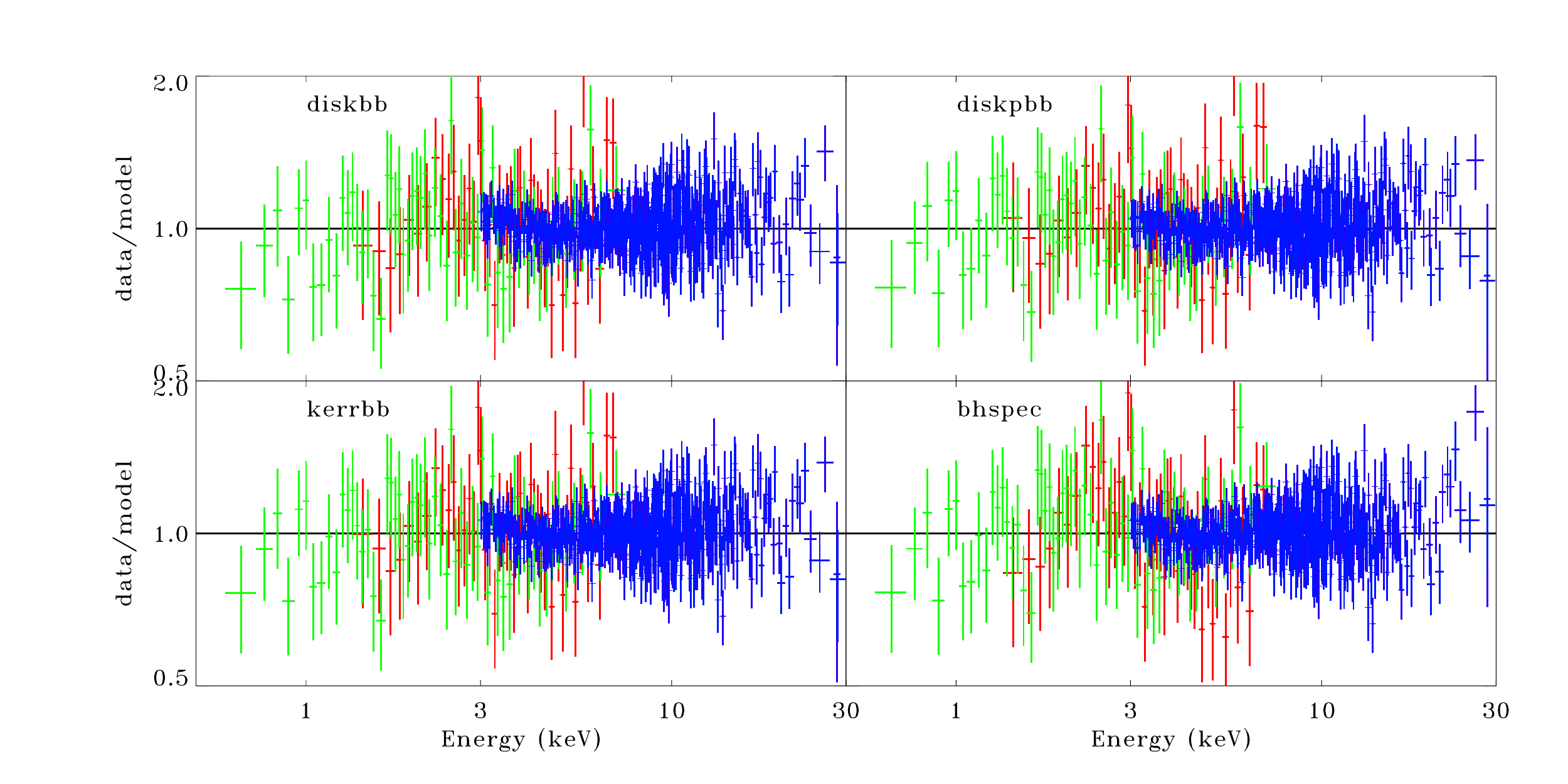}
\caption{Top - Combined \nustar\ (blue), \chandra\ (diffuse emission shown in purple, X-2 shown in red) and \swift\ (green) $EF_{\rm E}$ spectra, where different emission models for X-1 (shown with a solid black line) are fitted. Bottom - data-to-model ratios. For clarity, we do not show the ratios for the diffuse emission, which are shown in Figure \ref{fig_chandra_diffuse}.}
\label{fig_comb_spec2}
\end{center}
\end{figure*}

The main goal of this paper is to test various disk emission models for X-1 and to assess if the emission is consistent with the thin-disk model associated with sub-Eddington accreting black holes, or whether a significant departure from this is seen which could imply super-Eddington accretion. We begin by testing the simplest thin-disk model {\tt diskbb} (Figure \ref{fig_comb_spec2}, top left panel), which describes the emission from an accretion disk with multiple blackbody components with a temperature, $T$, that varies with the radius, $r$, as $T(r)\propto r^{-0.75}$. The parameters of the model are the temperature at the inner disk radius and the normalization, which is related to the inner disk radius as $((R_{\rm in}/$km$)/(D/10 $kpc$))^2 \times $cos$\theta$, where $D$ is the distance to the source and $\theta$ is the inclination angle of the disk. 

The fit with this model yields a statistically acceptable fit with \chisq/DOF=972.8/964, where the inner temperature of the disk is constrained to be $T_{\rm in}=1.89^{+ 0.10}_{- 0.09}$~keV. As noted by several previous authors, this temperature is hotter than expected for an accretion disk around an IMBH \citep[e.g.][]{strohmayer03}.

Following this, we progress to more sophisticated models, specifically those that include the black hole mass and spin as parameters and compute the Eddington ratio in a self-consistent manner. The first of these is {\tt kerrbb} \citep{li05}, which specifically takes into account general relativistic effects due to a spinning Kerr black hole (Figure \ref{fig_comb_spec2}, bottom left panel). The model accounts for self-irradiation of the disk and limb darkening, which can be switched on or off. The torque at the inner boundary of the disk is also allowed to be zero. We switch on self-irradiation and limb darkening and set the torque at the inner boundary to zero. The color correction factor, $\kappa$, which accounts for the deviation of the local disk spectrum from a black body due to electron scattering, is set to 1.7 \citep{davis11}. The fit with this model provides an improved fit over {\tt diskbb} with \chisq/DOF=961.1/962 and provides a constraint on the black hole mass of 80$^{+72}_{-60}$~\msol\ (where the spin is unconstrained between $a_{*}=0.99$ and $a_{*}=-0.99$) assuming that the disk extends to the ISCO. However, the model implies that M82~X-1 is shining at a super-Eddington rate, \lamedd=17$^{+106}_{-1}$.

{\tt bhspec} \citep{davis05, davis06} is similar to {\tt kerrbb} in that it is a fully relativistic accretion disk model, but differs from {\tt kerbb} by the treatment of the emission at the surface of the disk, where {\tt bhspec} does not assume the spectrum at the surface to be black body, but instead uses stellar-like atmosphere calculations to model the vertical structure of the disk (Figure \ref{fig_comb_spec2}, bottom right panel). We use a version of this model with the parameter ranges \mbh$=100-10^4$~\msol, \lamedd$=0.03-1$ and $a_{*}=0-0.99$. Since {\tt bhspec}, which is a grid model, does not allow for \lamedd$>1$, the results from this model are different from {\tt kerrbb}. A fit with this model gives \chisq/DOF=1002.6/962 and implies \mbh$=950^{+ 210}_{- 120}$~\msol, where the spin and Eddington ratios hit their upper limits, $a_{*}=0.9900^{+u}_{-0.0004}$ and \lamedd=1.00$^{+ u}_{- 0.01}$. By effectively assuming Eddington-limited accretion, the model is forced to higher black hole masses in order increase the flux, while simultaneously being forced to high spins to keep the temperature of the disk high. By being forced to its parameter limits, the results from {\tt bhspec} point towards super-Eddington accretion as does {\tt kerrbb}. However, at close-to-Eddington ratios and above, the standard thin disk that these models assume does not hold, thus we discard them as unphysical in this regime and do not rely on their black hole mass estimates.

Finally, the thermal emission from X-1 has previously been shown to be best described by a slim disk, rather than a thin disk \citep[e.g.][]{okajima06}. At high Eddington ratios, the standard thin disk is expected to transition to a slim disk which is dominated by advection \citep{abramowicz88} and has a different temperature profile as a function of radius than a thin disk does, where the temperature of the disk, $T$, at a given radius, $r$, is described as $T(r)\propto r^{-0.5}$ \citep{watarai00}. We test if M82~X-1 is consistent with the slim disk prediction using the simple {\tt diskpbb} model, which is similar to {\tt diskbb}, but with a variable exponent to the temperature profile, $p$ (i.e. $T(r)\propto r^{-p}$, Figure \ref{fig_comb_spec2}, top right panel). This model provides the best fit of all the disk models we have tested with \chisq/DOF=956.6/963. The derived parameters are $T_{\rm in}=2.36^{+ 0.29}_{- 0.30}$~keV and $p=0.55^{+ 0.07}_{- 0.04}$. Allowing a variable $p$ leads to a significant improvement in the fit statistic over a fixed $p$ of 0.75 (\dchisq$=-16$). The derived $p$ value is significantly lower than for a thin disk, and consistent with a slim disk, in confirmation of previous results (e.g. M09).

While the slim disk state indicated by the data imply a high-Eddington ratio, the {\tt diskpbb} model we use to model the spectrum is in fact a simplification since at high Eddington rates, outflows, geometric collimation and other effects are also expected to influence the observed emission. Nonetheless, radiation hydrodynamics simulations of super-Eddington accretion have been carried out \citep[e.g.][]{kawashima12,jiang14} and produce spectra that are roughly consistent with the spectral shape of slim-disk models. For example, \cite{kawashima12} compare their predicted spectra with the shape of the observed spectrum of several ULXs. They find that for face-on geometries, their results compare very well to the spectrum of NGC~1313~X-2, which is also well fitted with a {\tt diskpbb} model with $p\approx$0.5 \citep{gladstone09,bachetti13}.

A black hole mass estimate can also be made from slim-disk models, which is inferred from the inner radius measured by the model. To estimate the mass this way, we follow the prescription of \cite{soria15}, who estimated the mass of M83 ULX-1, which was also found to be in the slim-disk regime. The inner radius of the disk is related to the normalization of the model: $R_{\rm in}=\xi\kappa^2N^{1/2}($cos$\theta)^{-1/2}(D/10~{\rm kpc})$ km, where $\xi$ is a geometric factor and depends on how close to the ISCO the disk reaches its maximum temperature, $\kappa$ is the color correction factor, $N$ is the normalization of the model, and $\theta$ is the inclination of the disk. The mass of the black hole is related to $R_{\rm in}$ by \mbh$=1.2R_{\rm in}c^2/G\alpha$, where $\alpha=1.24$ for a maximally spinning black hole in an astrophysical context \citep[$a_*=0.998$,][]{thorne74} or $\alpha=6$ for a black hole with no spin. The factor of 1.2 results from the fact that the inner radius of a slim disk extends within the ISCO \citep{vierdayanti08}. While for standard thin-disks, $\xi=0.412$ \citep{kubota98} and $\kappa=1.7$ \citep{shimura95, davis05}, \cite{soria15} note that at high Eddington rates, $\kappa$ increases to 3 \citep[e.g.][]{watarai03} and $\xi=0.353$ \citep{vierdayanti08}, which we adopt here.

Given the normalization of the {\tt diskpbb} model, $N=0.033^{+0.028}_{-0.014}$, the distance of 3.3 Mpc to M82 and assuming a face-on inclination of the disk ($\theta=0$\degree), the formula above yields \mbh=$26^{+9}_{-6}$~\msol\ for a non-spinning black hole, or \mbh=$125^{+45}_{-30}$~\msol\ for a maximally spinning black hole. The luminosity of the disk from the {\tt diskpbb} model is 5.11$^{+ 0.42}_{- 0.30}\times10^{40}$~\ergs, which corresponds to \lamedd=14$^{+5}_{-3}$ for a non-spinning black hole, or \lamedd=3$\pm1$ for a maximally spinning black hole. 

The black hole mass estimates are also degenerate on the inclination angle assumed. Given an extreme inclination of 85\degree, the formula above yields \mbh=$88^{+30}_{-22}$~\msol\ for a non-spinning black hole, or \mbh=$424^{+149}_{-104}$~\msol\ for a maximally spinning black hole. which corresponds to \lamedd=0.9$^{+0.3}_{-0.3}$ for a non-spinning black hole, or \lamedd=3$\pm1$ for a maximally spinning black hole.

We list the best fit spectral parameters of the models above in Table \ref{table_specpar3} and show the fitted spectra and data-to-model ratios in Figure \ref{fig_comb_spec2}.

\begin{table*}
\centering
\caption{Spectral fitting results }
\label{table_specpar3}
\begin{center}
\begin{tabular}{l l l l l}
\hline
X-1					& {\tt diskbb}						& {\tt diskpbb}					& {\tt kerrbb}					& {\tt bhspec}	\\				
\hline
\nh\ ($10^{22}$ \cmsq) 	& 1.21$^{+ 0.33}_{- 0.26}$			& 2.07$^{+ 0.64}_{- 0.55}$ 		&  1.47$^{+ 0.27}_{- 0.24}$		&  1.42$^{+ 0.32}_{- 0.22}$ \\
$T_{\rm in}$ (keV) 		& 1.89$^{+ 0.10}_{- 0.09}$ 			& 2.36$^{+ 0.29}_{- 0.30}$ 	 	& -							& - \\
$p$ 					& -								& 0.55$^{+ 0.07}_{- 0.04}$ 	 	& -							& - \\
$a^*$ 				& -								& -							& 0.9999$^{+u}_{-l}$				& 0.9900$^{+u}_{-0.0004}$ \\
$i$ (\degree) 			& -								& -							& 47.4$^{+22.1}_{-26.2}$			& 84.3$^{+ 2.9}_{- 3.2}$ \\
$M_{\rm BH}$ (\msol)	& -								& -							&  80$^{+  72}_{-  60}$			&  955$^{+ 206}_{- 121}$\\
\lamedd				& -								& -							&  16.6$^{+106}_{-  0.9}$			&  1.00$^{+ u}_{- 0.01}$ \\
\fx\ ($10^{-11}$ \ergcms)	& 2.98$^{+ 0.14}_{- 0.10}$ 			& 3.19$^{+ 0.08}_{- 0.20}$ 	 	& 3.07$^{+ 0.15}_{- 0.13}$		&2.89$^{+ 0.14}_{- 0.11}$ \\
\lx\ ($10^{40}$ \ergs)		& 4.78$^{+ 0.26}_{- 0.12}$ 			& 5.11$^{+ 0.42}_{- 0.30}$		 	&5.27$^{+ 0.21}_{- 0.25}$			&4.96$^{+ 0.20}_{- 0.24}$ \\
\hline
X-2					&& && \\
\hline
\nh\ ($10^{22}$  \cmsq)	&2.21$^{+ 0.70}_{- 0.67}$ 			& 2.72$^{+ 0.95}_{- 0.79}$	 	&2.54$^{+ 0.86}_{- 0.79}$			&1.51$^{+0.22}_{-0.36}$ \\
$\Gamma$ 			& 0.38$^{+ 0.42}_{- 0.43}$			& 0.79$^{+ 0.58}_{- 0.52}$	 	&0.64$^{+ 0.52}_{- 0.54}$			&-0.35$^{+0.19}_{-0.18}$ \\
$E_{\rm C}$ (keV)		& 7.56$^{+ 3.09}_{- 1.66}$			& 11.22$^{+13.60}_{- 3.82}$	 	&9.35$^{+ 6.33}_{- 2.59}$			&4.88$^{+0.45}_{-0.36}$ \\
\fx\ ($10^{-11}$ \ergcms)	& 0.88$^{+ 0.05}_{- 0.14}$ 			& 0.69$^{+ 0.17}_{- 0.05}$	 	&0.78$^{+ 0.12}_{- 0.11}$			&1.01$^{+0.03}_{-0.03}$ \\
\lx\ ($10^{40}$ \ergs)		& 1.81$^{+ 0.08}_{- 0.08}$			& 1.74$^{+ 0.26}_{- 0.18}$		&1.75$^{+ 0.09}_{- 0.14}$			& 2.22$^{+ 0.06}_{- 0.05}$ \\
\hline
\chisq\ 				& 972.8 							& 956.6					 	&  961.1						& 1002.6 \\
DOF 				&964								&963							&962							&          962\\
\rchisq\ 				&1.009							&0.993						&0.999						& 1.042 \\
\hline
\end{tabular}
\tablecomments{Best-fit parameters for the disk models fitted to X-1 and the cutoff power-law for X-2. Fluxes are observed (not corrected for absorption) given in the 0.5$-$10~keV range, and the luminosities are total integrated over the whole model assuming a distance of 3.3 Mpc to M82 and corrected for absorption.}

\end{center}
\end{table*}

\section{Comparison with previous results}
\label{sec_comp}

As described above, we found that the spectrum of M82~X-1 requires a departure from the standard disk model, with a $p$-value of 0.55$^{+0.07}_{-0.04}$, which is significantly less than the $p=0.75$ that reproduces the standard disk, indicating that advection is significant in the disk. This was found similarly by M09, although with larger $p$-values (0.61-0.65). The best-fit temperature of the {\tt diskpbb} model is 2.36$^{+0.29}_{-0.30}$~keV, which is not as hot as what M09 find (3.4-3.6~keV). It should be noted that the luminosity of X-1 during our observations was a factor of $2-4$ higher than when it was observed by \suzaku, thus our results may not be strictly comparable. Interestingly, an increase in the disk temperature with decreasing luminosity is predicted by the model of supercritically accreting black holes of \cite{poutanen07}. However, M09 did not benefit from simultaneous spatially resolved \chandra\ data and they estimate the contribution from X-2 from previous \chandra\ observations, leading to systematic uncertainties in their best-fit parameters. Similarly, \cite{okajima06} fitted a long \xmm\ exposure of M82 with the {\tt diskpbb} model, which gave $p=0.61^{+0.03}_{-0.02}$ and a temperature of 3.73$^{+0.58}_{-0.40}$~keV. However, these authors neglected contributions to the \xmm\ spectrum by X-2 and the diffuse emission.

In addition to the {\tt diskpbb} model, \cite{okajima06} fit their \xmm\ spectrum of M82 with their own slim-disk model, deriving a black hole mass between 19$-$32~\msol, depending on the physical processes assumed (e.g. blackbody emission, Comptonisation, gravitational redshift, relativistic effects). This is consistent with our mass estimate of 26~\msol\ for a non-spinning black hole. \cite{okajima06} do not consider a spinning black hole.

\cite{feng10} also observed X-1 during a period of high flux with \chandra\ and \xmm. They fit the spectrum with the {\tt diskbb} model, obtaining disk temperatures from 1.10~keV to 1.52~keV, depending on the luminosity which ranged from $1.8-7.9\times10^{40}$~\ergs. Our fit with this model yielded a temperature of 1.89$^{+0.10}_{-0.09}$~keV and a disk luminosity of 4.78$^{+0.26}_{-0.12}\times10^{40}$~\ergs, a somewhat higher temperature than the range found by \cite{feng10}, despite a comparable luminosity. \cite{feng10} also applied the {\tt kerrbb} model to their combined \chandra\ and \xmm\ dataset in order to estimate the black hole mass. They test two different scenarios for the spin of the black hole, $a_{*}$=0 and $a_{*}$=0.9986. They rule out a non-rotating black hole due to the model yielding an Eddington ratio exceeding the limit by a factor of 160. Their maximally spinning scenario yields a black hole mass in the range 300$-$810~\msol\ with a disk inclination of 59$-$79 degrees (90\% confidence) and an Eddington ratio of 2.5.  These authors did not however consider a slim disk. Furthermore, they relied on a spectral model to account for pile-up, which introduces uncertainties into the spectral parameters and luminosity estimations, and performed their fits in a relatively narrow energy band with no coverage above 10~keV. Pile-up does not affect the \nustar\ data at the observed count rates.

Our results put the mass of M82~X-1 at the lower end of previous estimates, and are consistent with a stellar remnant black hole without spin, or a borderline IMBH with maximal spin. Interestingly the mass range is consistent with the masses of the merging black holes discovered through the detection of their gravitational wave signal \citep{abbott16}. Our mass estimates are considerably lower than those from thin-disk modeling \citep{feng10} or QPO analysis \citep{pasham14}. A subsequent analysis of the QPOs by \cite{stuchlik15} found that the black hole mass estimate from these are strongly model dependent, and given this put a range on the mass as $140<$\mbh$<660$~\msol, where the lower masses correspond to a non-spinning black hole, and as such are at odds with our result. In our estimation, we have assumed a face-on inclination of the disk. Larger inclinations would increase the mass estimate by (cos$\theta$)$^{-1/2}$, but even an extreme inclination value, such as $\theta=85$\degree, only increases the non-spinning estimate to \mbh=$62^{+22}_{-15}$~\msol.

\section{Alternative interpretations of the high-energy spectrum}
\label{sec_alt}

For all of the continuum models we have fitted for X-1, the spectrum turns down above 10~keV, however a significant signal remains, which we have modeled with a cutoff power-law representing emission from X-2. However, since the spatial resolution of \nustar\ does not allow us to determine directly that X-2 is indeed the source of the emission, we consider other scenarios for its origin. 

For other ULXs, an additional hard component has been found in excess the disk component. This has been interpreted as Compton-scattered disk emission by a corona \citep[e.g.][]{gladstone09,walton15}. We therefore test if the spectral shape above 10~keV, can be explained by such a component for X-1. For this we add the {\tt simpl} model \citep{steiner09}, which takes a fraction of the seed disk model photons (we test with our best-fitting {\tt diskpbb} model) and up-scatters them into a power-law component. The parameters of this model are the power-law index ($\Gamma$) and the scattered fraction ($f_{\rm scatt}$). 

We find that the addition of this component does not change the fit since the hard emission is already being accounted for by the {\tt cutoffpl} model for X-2. To test to what extent the {\tt simpl} model can account for this instead, we fix the parameters of the {\tt cutoffpl} model to the best-fit parameters found by the long \chandra\ exposure in \cite{brightman16} ($\Gamma=0.70$ and $E_{\rm C}=6.19$~keV), since these parameters lead to a softer spectrum for X-2. In this case the best fit {\tt simpl} parameters are $\Gamma=2.37^{+ 0.44}_{- 0.43}$ and $f_{\rm scatt}=0.15^{+ 0.09}_{- 0.04}$ with \chisq/DOF=974.7/963. In comparison, for Holmberg II X-1 \cite{walton15} found $\Gamma=3.1^{+ 0.3}_{- 1.2}$ and $f_{\rm scatt}=0.4^{+ 0.5}_{- 0.3}$, which are similar to what we find for M82 X-1 considering the uncertainties.

The addition of the {\tt simpl} model also leads to changes in the {\tt diskpbb} model parameters which become $T_{\rm in}=2.01^{+ 0.13}_{- 0.19}$~keV, $p=0.59^{+ 0.06}_{- 0.04}$ and $N=0.075^{+0.028}_{-0.014}$. The temperature and $p$-value are within the statistical uncertainties of the previous fit with {\tt diskpbb}, however the normalization has increased. This in turn leads to an increase in the black hole mass estimate, which given the same assumptions as above becomes \mbh=$39^{+14}_{-8}$~\msol\ for a non-spinning black hole, or \mbh=$188^{+70}_{-38}$~\msol\ for a maximally spinning black hole. These estimates are not significantly different from our initial estimates. 

Another alternative source of the high-energy signal is the hot diffuse gas in the center of M82 in which the ULXs are embedded. We have modeled this with three {\tt apec} models with temperatures of 0.28, 1.01 and 7~keV. For the first two components, the temperatures were constrained by the \chandra\ data, however the highest temperature component could not be constrained, and was fixed at 7~keV based on previous results by \cite{ranalli08}. We now explore if this hot {\tt apec} component can account for the high-energy signal by freeing the temperature. For this we keep the spectrum of X-2 fixed as described above. We find that the signal can indeed be described by a hot {\tt apec} model where $kT=13.2^{+2.4}_{-1.0}$~keV with \chisq/DOF=963.31/962. While \cite{ranalli08} find that their temperature profile peaks at $\sim7$ keV, their data and/or model do not allow for exploration of temperatures $>10$ keV, thus the temperature that we found may be consistent with their results. We note that \cite{cappi99} also found evidence for a hot thermal component in M82 from {\it BeppoSAX} data, with a temperature of $\sim5-8$~keV. The origin of this emission was discussed by \cite{ranalli08}, including non-thermal bremsstrahlung emission from star formation and unresolved point sources, however these were ruled out and no clear conclusion was reached.

For this scenario, the parameters of the {\tt diskpbb} model for X-1 become $T_{\rm in}=2.21^{+ 0.20}_{- 0.15}$~keV, $p=0.55^{+ 0.06}_{- 0.04}$ and $N=0.042^{+0.032}_{-0.017}$. These new parameters imply \mbh=$29^{+10}_{-7}$~\msol\ for a non-spinning black hole, or \mbh=$140^{+46}_{-32}$~\msol\ for a maximally spinning black hole. Again these parameters are very similar to those assuming that the high-energy emission originates from X-2, and therefore the assumption regarding the origin of this emission does not significantly change our results regarding X-1.

\section{Summary and Conclusions}
\label{sec_conc}

In this paper we have presented analysis of simultaneous \nustar, \chandra\ and \swiftxrt\ observations of the ultraluminous X-ray source M82~X-1 during a period of flaring activity. The \chandra\ data have allowed us to spatially resolve the source from the other bright sources of X-rays in the galaxy, specifically that of the nearby ultraluminous pulsar, X-2, and the bright diffuse emission. Combined with \nustar\ and \swiftxrt\ data, this provides a sensitive measurement of the 0.5$-$30~keV spectrum of the source. We have fitted standard thin accretion disk models for sub-Eddington accretion to the spectrum finding that they require super-Eddington accretion rates in order to reproduce the observed spectrum. Since the thin accretion disk models do not hold at high Eddington ratios, we discard the thin-disk models as unphysical. We directly test for the departure from the thin-disk model using a disk model that allows for a variable temperature as a function of radius of the disk ({\tt diskpbb}), finding that the temperature  profile is ($T(r)\propto r^{-0.55}$), which is significantly flatter than expected for a thin disk, and is instead characteristic of a slim disk, which is expected at high Eddington ratios. While at high Eddington rates, outflows and geometric collimation are also expected to influence the observed emission, which our simple model does not account for, radiation hydrodynamics simulations of super-Eddington accretion have shown that the predicted spectra are very similar to what we observe for M82~X-1. We therefore conclude that the ULX is a super-Eddington accretor. Our mass estimates inferred from the inner disk radius imply a stellar-remnant black hole (\mbh=$26^{+9}_{-6}$~\msol) when assuming zero spin, or an IMBH (\mbh=$125^{+45}_{-30}$~\msol) when assuming maximal spin.

\section{Acknowledgements}

This work made use of Director's Discretionary Time on \chandra, for which we thank Belinda Wilkes for approving and the \chandra\ X-ray Center for implementing. We also use Director's Discretionary Time on \nustar, for which we thank Fiona Harrison for approving and the \nustar\ SOC for co-ordinating with \chandra. The \nustar\ mission is a project led by the California Institute of Technology, managed by the Jet Propulsion Laboratory, and funded by NASA. We thank the \nustar\ Operations, Software and Calibration teams for support with the execution and analysis of these observations. This research has made use of the \nustar\ Data Analysis Software (NuSTARDAS) jointly developed by the ASI Science Data Center (ASDC, Italy) and the California Institute of Technology (USA). AZ acknowledges funding from the European Research Council under the European Union's Seventh Framework Programme (FP/2007-2013)/ERC Grant Agreement n. 617001.

{\it Facilities:} \facility{\chandra\ (ACIS), \nustar, \swift\ (XRT)}

\bibliography{bibdesk.bib}

\end{document}